\documentclass[apjl]{emulateapj}
\begin{document}

\title{New Constraints on the Galactic Bar}

\author{I.~Minchev\altaffilmark{1}, J.~Nordhaus\altaffilmark{1,2} 
and A.~C.~Quillen\altaffilmark{1}}

\altaffiltext{1}{Department of Physics and Astronomy, University of Rochester, 
Rochester, NY 14627; iminchev@pas.rochester.edu, aquillen@pas.rochester.edu}
\altaffiltext{2}{Laboratory for Laser Energetics, University of Rochester, 
Rochester, NY 14623; nordhaus@pas.rochester.edu}

\begin{abstract}
Previous work has related the Galactic Bar to structure in the local 
stellar velocity distribution. Here we show that the Bar also influences
the spatial gradients of the velocity vector via the Oort constants.
By numerical integration of test-particles we simulate measurements of 
the Oort $C$ value in a gravitational potential 
including the Galactic Bar. We account for the observed trend that $C$ is 
increasingly negative for stars with higher velocity dispersion.
By comparing measurements of $C$ with our simulations we 
improve on previous models of the Bar, estimating that the Bar pattern 
speed is $\Omega_{\rm b}/\Omega_0=1.87\pm0.04$, where $\Omega_0$ is the
local circular frequency, and the Bar angle lies within 
$20^\circ\leqslant\phi_0\leqslant45^\circ$. 
We find that the Galactic Bar affects measurements of the Oort constants 
$A$ and $B$ less than $\sim2$ km/s/kpc for the hot stars. 
\end{abstract}


\section{Introduction}
The Galaxy is often modeled as an axisymmetric disk.
With the ever increasing proper motion and radial velocity data, 
it is now apparent that nonaxisymmetric effects cannot be neglected
(nonzero Oort constant $C$, \citealt{olling03}, hereafter O\&D; 
nonzero vertex deviation, \citealt{famaey05};
asymmetries in the local velocity distribution of stars, \citealt{dehnen98}). 
It is still not clear, however, what the exact nature of the perturbing 
agent(s) in the Solar neighborhood (SN) is(are).
Possible candidates are spiral density waves, a central Bar, and a triaxial
halo.

Due to our position in the Galactic disk, the properties of the Milky 
Way Bar are hard to observe directly. 
Hence its parameters, such as orientation and pattern speed, have
only been inferred indirectly from observations of the inner Galaxy 
(e.g., \citealt{blitz91,weinberg92}). 

However, the Bar has also been found to affect the local velocity distribution 
of stars. {\it HIPPARCOS} data revealed more clearly
a stream of old disk stars with an asymmetric drift of about 45 km/s and 
a radial velocity $u<0$, with $u$ and $v$ positive toward the Galactic 
center and in the direction of Galactic rotation, respectively.
This agglomeration of stars has been dubbed the ``Hercules" stream or the
``$u$-anomaly". The numerical work of \cite{dehnen99,dehnen00} and 
\cite{fux01} has shown that this stream can be explained as the effect 
of the Milky Way Bar if the Sun is placed just outside the 2:1 Outer Lindblad
Resonance (OLR). Using high-resolution spectra of nearby F and G 
dwarf stars, \cite{bensby07} have investigated the detailed
abundance and age structure of the Hercules stream. Since this stream 
is composed of stars of different ages and metallicities pertinent to
both thin and thick disks, they concluded that a dynamical effect, such
as the influence of the Bar, is a more likely explanation than a dispersed 
cluster.

Assuming the Galactic Bar affects the shape of the distribution function
of the old stellar population in the SN, an additional 
constraint on the Bar can be provided by considering the values derived 
for the Oort constant $C$. In other words, in addition to relating the 
dynamical influence of the Bar to the local velocity field, $C$ provides
a link to the gradients of the velocities as well.
The study of O\&D not only measured a non-zero $C$, implying the presence 
of non-circular motion in the SN, but also found that $C$ 
is  more negative for older and redder stars with a larger velocity 
dispersion. This is surprising as a hotter stellar population
is expected to have averaged properties more nearly axisymmetric
and hence, a reduced value of $|C|$ (e.g., \citealt{mq07}; hereafter Paper I).

It is the aim of this letter to show it is indeed possible to explain the
observationally deduced trend for $C$ (O\&D) 
by modeling the Milky Way as an exponential disk perturbed by a central Bar. 
By performing this exercise, we provide additional constraints on the Bar's 
pattern speed. Models for the structure in the bulge of our galaxy are 
difficult to constrain because of the large numbers of degrees of freedom 
in Bar models. Thus future studies of structure in the Galactic Center 
will benefit from tighter constraints on the parameters describing the Bar, 
such as its pattern speed and angle with respect to the Sun.

\section{The Oort constants}

We can linearize the local velocity field (e.g., Paper I) about the
LSR and write the mean radial velocity $\overline{v}_d$ and longitudinal proper 
motion $\overline{\mu}_l$ as functions of the Galactic longitude $l$ as
\begin{eqnarray}
\label{eq:vel}
{\overline{v}_d\over \overline{d}} &=& K + A\sin(2l) + C\cos(2l)   \\
\overline{\mu}_l &=& B + A\cos(2l) - C\sin(2l)\nonumber
\end{eqnarray}
where $\overline{d}$ is the average heliocentric distance of stars, $A$ and $B$ are
the usual Oort constants, and $C$ and $K$ are given by
\begin{eqnarray}
\label{eq:oc}
2C  &\equiv &   -{\overline{u} \over r} + {\partial \overline{u} \over \partial r}
- {1\over r} {\partial \overline{v}_{\phi} \over \partial \phi}         \\
2K  &\equiv &   +{\overline{u} \over r} + {\partial \overline{u} \over \partial r}
+ {1\over r} {\partial \overline{v}_{\phi} \over \partial \phi}.
\end{eqnarray}
Here $r$ and $\phi$ are the usual polar coordinates and $v_\phi=v_0+v$, where 
$v_0$ is the circular velocity at the Solar radius, $r_0$. In this work we primarily 
consider a flat rotation curve (RC) (see, however \S \ref{sec:concl}), hence the 
derivatives of $v_\phi$ in the above equations are identical to the derivatives of 
$v$. $C$ describes the radial shear of the velocity field and $K$ its divergence.
For an axisymmetric Galaxy we expect vanishing values for both
$C$ and $K$ \footnote{Note, however, that $C$ and $K$ would also be zero 
in the presence of nonaxisymmetric structure if the Sun happened to be 
located on a symmetry axis.}.
Whereas $C$ could be derived from both radial velocities and
proper motions, $K$ can only be measured from radial velocities, in which 
case accurate distances are also needed.

A problem with using proper motions data has been described by 
O\&D. The authors
present an effect which arises from the longitudinal variations of the 
mean stellar parallax caused by intrinsic density inhomogeneities. Together
with the reflex of the solar motion these variations create contributions
to the longitudinal proper motions which are indistinguishable from the
Oort constants at $\le$ 20\% of their amplitude. O\&D corrected 
for the ``mode-mixing" effect described above, using the latitudinal proper 
motions. 
The resulting $C$ is found to vary approximately 
linearly with both color and asymmetric drift (and thus mean age) from
$C\approx0$ km/s/kpc for blue samples to $C\approx-10$ km/s/kpc for 
late-type stars (see Figs. 6 and 9 in O\&D). 
Since $C$ is related to the radial streaming of stars, 
we expect non-axisymmetric structure to mainly affect the low-dispersion
population which would result in the opposite behavior for $C$. 
In Paper I we showed that spiral structure failed to explain the
observed trend of $C$.

Note that the Oort constants are not constant unless they are 
measured in the SN. Due to nonaxisymmetries 
they may vary with the position in the Galaxy. 
Thus, the Oort constants have often been called the Oort {\it functions}.

\section{The simulations}

\begin{deluxetable}{lcc}
\tablewidth{0pt}
\tablecaption{Simulation parameters used\label{table:par}}
\tablehead{
\colhead{Parameter}                  &
\colhead{Symbol}                    &
\colhead{Value}           
} 
\startdata
Solar neighborhood radius   & $r_0$  &   1                            \\
Circular velocity at $r_0$  & $v_0$  &   1                            \\
Radial velocity dispersion & $\sigma_{\rm u}(r_0)$ & $0.05v_0$ or $0.18v_0$ \\
$\sigma_{\rm u}$ scale length & $r_{\sigma}$ & $0.9r_0$                     \\
Disk scale length &  $r_\rho$       &    $0.37r_0$                    \\
Bar strength      &  $\epsilon_{\rm b}$   &     $-0.012$              \\
Bar size          &   $r_{\rm b}$         &     $0.8r_{\rm cr}$
\enddata 
\end{deluxetable}

\begin{figure}[t!]
\resizebox{\hsize}{!}{\includegraphics{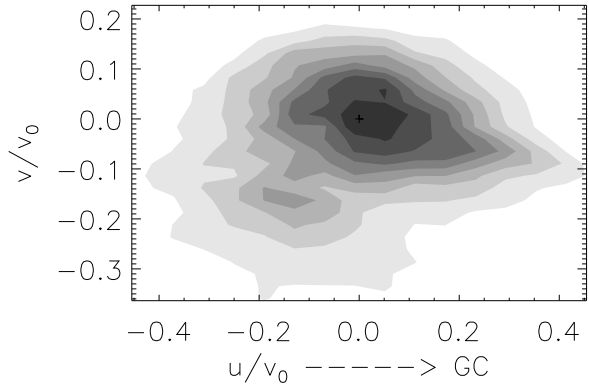}}
\caption{
Simulated $u-v$ distribution for a Bar pattern speed $\Omega_{\rm b}=1.87\Omega_0$,
Bar angle $\phi_0=35^\circ$ and a sample depth $d=r_0/40$. The initial velocity
dispersion is $\sigma_{\rm u}=0.18v_0$ and the Bar strength is $\epsilon_{\rm b}=-0.012$.
Contour levels are equally spaced. The clump identified with
the Hercules stream is clearly discernible in the lower left portion
of the plot. This figure is in agreement with the observed $f_0(u,v)$
\citep{dehnen98,fux01} and simulated distribution functions of
\cite{dehnen00} and \cite{fux01}.
}
\label{fig:uv}
\end{figure}

We perform 2D test-particle simulations of an initially axisymmetric 
exponential galactic disk. To reproduce the observed kinematics of the Milky Way, 
we use disk parameters consistent with observations (see Table \ref{table:par}). 
A detailed description of our disk model and simulation technique can be found 
in Paper I. We are interested in the variation of $C$ with color ($\bv$), and 
asymmetric drift $v_a$. We simulate variations with color by assuming
that the velocity dispersion increases from blue to red star samples.
For a cold disk we start with an initial radial velocity dispersion 
$\sigma_{\rm u}=0.05v_0$ whereas for a hot disk we use $\sigma_{\rm u}=0.18v_0$.
The background axisymmetric potential due to the disk and halo has the form 
$\Phi_0(r)=v_0^2\log(r)$, corresponding to a flat RC.

\subsection{The Bar potential}

\begin{figure*}
\epsscale{1.2}
\plotone{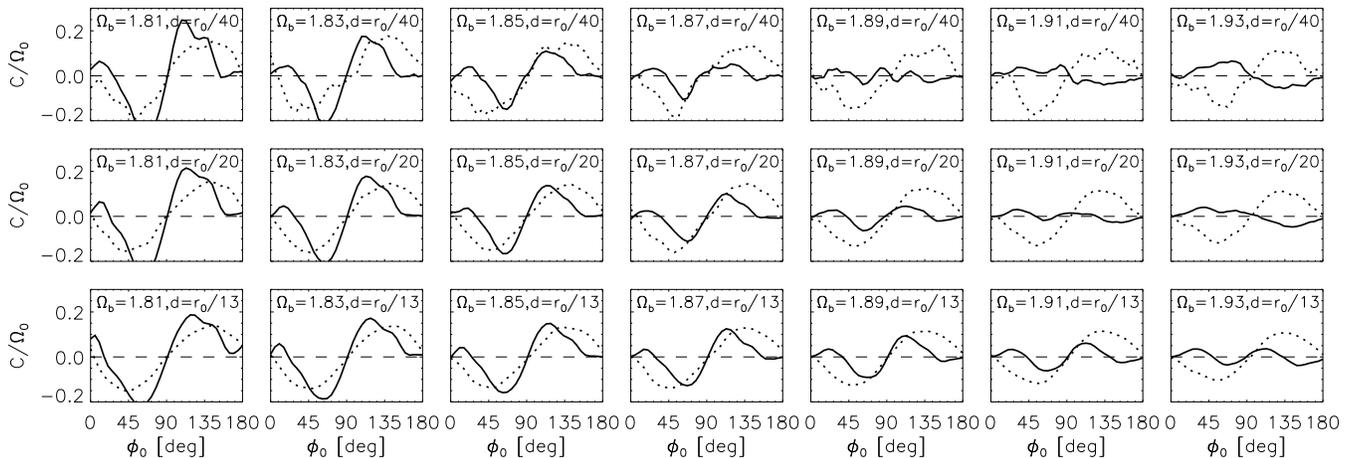}
\figcaption{
Each panel shows the variation of the Oort constant $C$ with Bar angle $\phi_0$,
for a simulation with the parameters given in Table \ref{table:par} and a particular
Bar pattern speed, $\Omega_{\rm b}$, and a mean sample depth, $\overline{d}$.
Solid and dotted lines correspond to cold- and hot-disk values, respectively.
Columns from left to right, show an increasing $\Omega_{\rm b}$ in units of
$\Omega_0$. Note that the OLR is at $\Omega_{\rm OLR}\approx1.7$. Different rows
present results from samples with different mean heliocentric distance, $\overline{d}$.
Good matches to the observed trend in $C$ (vanishing value for the cold disk and
a large negative for the hot one) are achieved for
$20^\circ\leqslant\phi_0\leqslant45^\circ$ and
$1.83\leqslant\Omega_{\rm b}/\Omega_0\leqslant1.91$.
}
\label{fig:c}
\end{figure*}

We model the nonaxisymmetric potential perturbation due to the Galactic Bar 
as a pure quadrupole
\begin{equation}
\Phi_{\rm b} = A_{\rm b}(\epsilon_{\rm b}) \cos[2(\phi-\Omega_{\rm b} t)]\times\left\{
\begin{array}{cclcr}
         \left(r_{\rm b}\over r\right)^3  &,&  r&\ge & r_{\rm b}    \\ 
       2-\left(r\over r_{\rm b}\right)^3  &,&  r&\le & r_{\rm b}
\end{array}
\right.
\end{equation}
Here $A_{\rm b}(\epsilon_{\rm b})$ is the Bar's gravitational potential 
amplitude, identical to the same name parameter used by \cite{dehnen00}; 
the strength is specified by $\epsilon_{\rm b}=-\alpha$
from the same paper. 
The Bar length is $r_{\rm b}=0.8r_{\rm cr}$ with $r_{\rm cr}$ the Bar corotation
radius.  The pattern speed, $\Omega_{\rm b}$ is kept constant. 
We grow the Bar by linearly varying its 
amplitude, $\epsilon_{\rm b}$, from zero to its maximum value in four Bar 
rotation periods.
We present our results by changing $\Omega_{\rm b}$ and keeping $r_0$ fixed.
The 2:1 outer Lindblad resonance (OLR) with the Bar is achieved when  
$\Omega_{\rm b}/\Omega_0=1+\kappa/2\approx1.7$, where $\kappa$ is the epicyclic 
frequency. We examine a region of parameter space for a range of pattern 
speeds placing the SN just outside the OLR. For a given 
pattern speed one could obtain the ratio $r_0/r_{\rm OLR}$ through 
$r_0/r_{\rm OLR}=\Omega_{\rm b}/\Omega_{\rm OLR}\approx\Omega_{\rm b}/1.7$. 
 
In contrast to \citet{dehnen00} and similar to \citet{fux01} and 
\citet{muhlbauer03}, we integrate forward in time. 
To allow for phase mixing to complete after the Bar reaches its maximum 
strength, we wait for 10 Bar rotations before we start recording the 
position and velocity vectors. In order to improve statistics, positions 
and velocities are time averaged for 10 Bar periods. 
After utilizing the two-fold symmetry of our galaxy we end up with
$\sim5\times10^5$ particles in a given simulated solar neighborhood with 
maximum radius $d_{\rm{max}}=r_0/10$. 

\section{Results}

First we show that we can reproduce the results of
Dehnen and Fux for the Hercules stream. We present a simulated $u-v$ 
velocity distribution for $\Omega_{\rm b}=1.87\Omega_0$, $\phi_0=35^\circ$, and a 
sample depth $\overline{d}=r_0/40$ in fig. \ref{fig:uv}. 
The initial velocity dispersion is $\sigma_{\rm u}=0.18v_0$.
This plot is indeed in very good agreement with previous 
test-particle \citep{dehnen00,fux01} and N-body \citep{fux01} simulations.

By a quantitative comparison of the observed with the simulated distributions,
\cite{dehnen00} deduced the Milky Way Bar pattern speed to be 
$\Omega_{\rm b}/\Omega_0=1.85\pm0.15$. In this calculation only the local velocity
field, i.e., the observed $u-v$ distribution, was taken into account.
Now, if one could also relate the dynamical effect of the Bar to the 
derivatives of the velocities, an additional constraint on Bar parameters
would be provided. 
Since the gradients of $u$ and $v$ are hard to measure directly, one needs
an indirect way of achieving this task. The obvious candidates, as anticipated, 
are the Oort constants $C$ and $K$. Oort's $K$, however, is hard to measure as 
mentioned above; hence only $C$, as estimated by O\&D, will be employed here.
By Fourier expansion of eqs. \ref{eq:vel} (see Paper I for details) we 
estimate $C$ from our numerically simulated velocity distributions
for the initially cold and hot disks.

\subsection{Variation of $C$ with Bar pattern speed and orientation}
\label{sec:c}
 
In Fig. \ref{fig:c} we present our results for $C$ as a function of the Bar angle,
$\phi_0$ (the angle by which the Sun's azimuth lags the Bar's major axis). 
Each column shows a simulation with a different pattern speed, indicated
in each plot. Rows from top to bottom show $C$ as calculated from samples at 
average heliocentric distances corresponding to $\overline{d}=200, \overline{d}=400$, 
and $\overline{d}=600$ pc, for a Solar radius $r_0=7.8$ kpc. Solid and dotted 
lines represent the results for cold and hot disks, respectively. The dashed lines 
indicates $C=0$. $C$ is presented in units of $\Omega_0=v_0/r_0$. To make the 
discussion less cumbersome, we write $C_{\rm h}$ and $C_{\rm c}$ to refer to 
the values for $C$ as estimated from the hot and cold disks, respectively.  

$C_{\rm h}$ (dotted lines in Fig. \ref{fig:c}) varies with galactic azimuth as 
$C_{\rm h}(\phi_0)\sim\sin(2\phi_0)$ for all of the $\Omega_{\rm b}$ values 
considered. On the other hand, the cold disk values (solid line) exhibit different
variations, depending on $\Omega_{\rm b}$ or equivalently, on the ratio 
$r_0/r_{OLR}$.  
Closer to the OLR (left columns of Fig. \ref{fig:c}), $C_{\rm c}(\phi_0)$ 
approaches the functional behavior of $C_{\rm h}(\phi_0)$. Away from the OLR 
$C_{\rm c}(\phi_0)$ is shifted by $90^\circ$ compared to $C_{\rm h}(\phi_0)$.
While both cold and hot disks yield in increase in the amplitude of $C(\phi_0)$ as 
the pattern speed nears the OLR, the effect on the $C_{\rm c}(\phi_0)$ is much
stronger.
This is consistent with our expectation that the cold disk is affected more
by the Bar, especially near the OLR. While close to the OLR 
$|C_{\rm h}(\phi_0)|<|C_{\rm c}(\phi_0)|$, we observe the opposite behavior away from 
it. This could be explained by the results of \cite{muhlbauer03}, where they find
that high velocity dispersion stars tend to shift the ``effective
resonance" radially outwards.

According to O\&D's ``mode-mixing" 
corrected value, $C\approx0$ for the cold sample and decreases to about -10 
km/s for the hot population. Hence we look for locations in Fig. \ref{fig:c}
complying with this requirement. 
In the transition region, $1.83\leqslant\Omega_{\rm b}/\Omega_0\leqslant1.91$, 
where the function $C_{\rm c}(\phi_0)$ transforms into its negative, there are
specific angles which provide good matches for the observations, i.e., 
$C_{\rm c}(\phi_0)\approx0$ while $C_{\rm h}(\phi_0)$ is significantly negative.
We performed simulations with different pattern speeds in the range 
$1.7\leqslant\Omega_b/\Omega_0\leqslant2.5$. It was found that only in the 
transition region in Fig. \ref{fig:c} could one achieve a satisfactory match to the 
observations. Although the left- and right-most columns are nearly consistent 
with our requirement on $C$, we reject these values of $\Omega_{\rm b}$ for the 
following reason. O\&D estimated a large negative $C$ from both the reddest 
main-sequence stars, which are
nearby, and the more distant red giants. This is inconsistent with the largest 
mean sample depth shown in Fig. \ref{fig:c} with $\Omega_{\rm b}=1.81,1.93$.
We conclude that the Bar pattern speed must lie in the range
\begin{equation}
\Omega_{\rm b}/\Omega_0=1.87\pm0.04.
\end{equation}

The Bar angle with respect to the Sun has been
proposed, as derived from IR photometry, to lie in the range
$15^\circ-45^\circ$ with the Sun lagging the Bar major axis, whereas 
\cite{dehnen00} found that the Bar reproduced the $u$-anomaly for 
$10^\circ\leqslant\phi_0\leqslant70^\circ$.
Examining Fig. \ref{fig:c}, we find that our requirements on $C$ result in 
constraining the Bar angle in the range
$20^\circ\leqslant\phi_0\leqslant45^\circ$, depending on $\Omega_{\rm b}$. 
This result for the dependence of pattern speed on Bar angle is in a very good 
agreement with Fig. 10 by \cite{dehnen00} which shows a linear increase of the 
derived $\Omega_{\rm b}/\Omega_0$ as a function of $\phi_0$. This is remarkable 
since \cite{dehnen00} obtained his results in a completely different way than
our method here.

\subsection{The effect of the Bar on $A$ and $B$}

The Oort constants $A$ and $B$ were found to also be affected by the Bar, although
to a much smaller extent. One could estimate the Bar-induced errors as
$\Delta A\equiv A-A_{\rm axi}$ and $\Delta B\equiv B-B_{\rm axi}$. Here
$A_{\rm axi}$, $B_{\rm axi}$ are the resulting axisymmetric values when
simulations excluding the Bar perturbation are performed. Note that we 
require a different simulation with each of the initial radial velocity dispersions 
used in order to estimate the asymmetric drift induced errors.
We found that for the cold disk the Bar causes $\Delta A=3.5\pm0.7$ 
and $\Delta B=3\pm0.6$ km/s/kpc, 
while the hot disk yielded $\Delta A=0.5\pm1.5$ and $\Delta B=0\pm1$ km/s/kpc. 
These values are calculated for $r_0=7.8$ kpc and $v_0=220$ km/s.
Despite the Bar's effect on $C$, the hot population does provide better
measurements for $A$ and $B$.

O\&D found an approximately linear variation of $A$ and $B$ with color with
$A$ increasing and $B$ decreasing. The Bar induced errors presented here
match the observed trend of $B$ but have the opposite one for $A$. It is possible
that local spiral structure is responsible for this discrepancy.

\section{Conclusion}
\label{sec:concl}

We have shown that the Galactic Bar can account for a trend seen in measurements 
of Oort's $C$, namely a more negative $C$-value with increasing velocity 
dispersion. This trend is difficult to explain with other dynamical models 
(e.g., Paper I). By requiring that the Bar model accounts for
this trend in $C$, we have improved the measurement of the Bar pattern 
speed, finding $\Omega_{\rm b}/\Omega_0=1.87\pm0.04$. In addition, the Bar angle
is found to be in the range $20^\circ\leqslant\phi_0\leqslant45^\circ$. 
Our result for $\Omega_{\rm b}$ lies well within the estimate by \cite{dehnen00}
and that by \cite{debattista02}, based on OH/IR star kinematics.
This study provides an improvement on the measurement of the Bar pattern speed
by a factor of $\sim4$ compared to previous work \citep{dehnen00}.
The improved constraints on Bar parameters should be tested by, and will 
benefit, future studies of the Bar structure in the Galactic center region, that
will become possible with future radial velocity and proper motion studies 
(e.g., BRAVA, \citealt{rich07}).

In addition to a flat RC we have also considered a power law initial
tangential velocity $v_\phi=v_0(r/r_0)^\beta$ with $\beta=0.1,-0.1$ corresponding
to a rising and a declining RC, respectively. We found that for
$\beta=0.1$ no additional error in $\Omega_b$ is introduced. However, for
$\beta=-0.1$ we estimated $\Omega_{\rm b}/\Omega_0=1.85\pm0.06$.
From Fig. \ref{fig:c} we see that the main source of error in the Bar pattern 
speed is due to the uncertainty in the Bar angle. A recent work by 
\cite{rattenbury07} used OGLE-II microlensing observations of red clump giants 
in the Galactic bulge, to estimate a bar angle of $24^\circ-27^\circ$.
Using this as an additional constraint we obtain
$\Omega_{\rm b}/\Omega_0=1.84\pm0.01$.

While we account for the trend of increasingly negative $C$ with increasing 
velocity dispersion, we fail to reproduce the size of the $C$-value measured by 
O\&D. The most negative value we predict is about -6 while they measure 
-10 km/s/kpc. We discuss possible reasons for this discrepancy:
(i) we expect that the magnitude of $C$ is related to the fraction of SN
stars composing the Hercules stream. It is possible that during its formation the 
Bar changed its pattern speed forcing more stars to be trapped in the 2:1
OLR, thus increasing this fraction. This would give rise to a more negative 
$C$ from hot stars, while leaving the cold population unaffected;
(ii) the numerical model here uses a distance limited sample. Modeling a stellar
population with a magnitude limited sample is more appropriate
to compare to the observed measurements.
(iii) O\&D's ``mode-mixing" correction is only valid if the Sun's motion in the 
$z$-direction were assumed constant. This is invalidated if, for example,
a local Galactic warp were present.

\acknowledgements
Support for this work was in part provided by NSF grant ASST-0406823, and 
NASA grant No.~NNG04GM12G issued through the Origins of Solar Systems Program.

{}


\begin{thebibliography}{}

\bibitem[Bensby et al.(2007)]{bensby07}
Bensby, T., Oey, M.~S.,
Feltzing, S., \& Gustafsson, B.\ 2007, \apjl, 655, L89

\bibitem[Blitz \& Spergel(1991)]{blitz91} 
Blitz, L., \& Spergel, D.~N.\ 1991, \apj, 379, 631

\bibitem[Debattista et al.(2002)]{debattista02} 
Debattista, V.~P., Gerhard, O., \& Sevenster, M.~N.\ 2002, \mnras, 334, 355

\bibitem[Dehnen(1998)]{dehnen98}
Dehnen, W.\ 1998, AJ, 115, 2384

\bibitem[Dehnen(1999)]{dehnen99} 
Dehnen, W.\ 1999, \apjl, 524, L35

\bibitem[Dehnen(2000)]{dehnen00} Dehnen, W.\ 2000, \aj, 119, 800 

\bibitem[Famaey et al.(2005)]{famaey05} 
Famaey, B., Jorissen, A., Luri, X., Mayor, M., Udry, S., 
Dejonghe, H., \& Turon, C.\ 2005, \aap, 430, 165

\bibitem[Fux(2001)]{fux01} 
Fux, R.\ 2001, \aap, 373, 511

\bibitem[Kuijken \& Tremaine(1991)]{kuijken91} 
Kuijken, K., \& Tremaine, S.\ 1991, in Sundelius B., ed.,
Dynamics of Disc Galaxies, Göteborg Univ. Press, p. 71

\bibitem[Minchev \& Quillen(2006)]{mq06} 
Minchev, I., \& Quillen, A.~C.\ 2006, \mnras, 368, 623

\bibitem[Minchev \& Quillen(2007)]{mq07} Minchev, I., \&
Quillen, A.~C.\ 2007, \mnras, 377, 1163 (Paper I)

\bibitem[M{\"u}hlbauer \& Dehnen(2003)]{muhlbauer03} 
M{\"u}hlbauer, G., \& Dehnen, W.\ 2003, \aap, 401, 975 

\bibitem[Olling \& Dehnen(2003)]{olling03} 
Olling, R.~P., \& Dehnen, W.\ 2003, \apj, 599, 275 (O\&D)

\bibitem[Rattenbury et al.(2007)]{rattenbury07}
Rattenbury, N.~J., Mao, S., Sumi, T., \& Smith, M.~C.\ 2007,
ArXiv e-prints, 704, arXiv:0704.1614

\bibitem[Rich et al.(2007)]{rich07} 
Rich, R.~M., Reitzel, D.~B., Howard, C.~D., \& Zhao, H.\ 2007, \apjl, 658, L29 

\bibitem[Torra et al.(2000)]{torra00} 
Torra, J., Fern{\'a}ndez, D., \& Figueras, F.\ 2000, \aap, 359, 82

\bibitem[Weinberg(1992)]{weinberg92} 
Weinberg, M.~D.\ 1992, \apj, 384, 81

\end{thebibliography}
\end{document}